\documentclass[sn-mathphys-num]{sn-jnl}% Math and Physical Sciences Numbered Reference Style 
\usepackage{graphicx}%
\usepackage{multirow}%
\usepackage{amsmath,amssymb,amsfonts}%
\usepackage{amsthm}%
\usepackage{mathrsfs}%
\usepackage[title]{appendix}%
\usepackage{xcolor}%
\usepackage{textcomp}%
\usepackage{manyfoot}%
\usepackage{booktabs}%
\usepackage{algorithm}%
\usepackage{algorithmicx}%
\usepackage{algpseudocode}%
\usepackage{listings}%
\theoremstyle{thmstyleone}%
\theoremstyle{thmstyletwo}%

\theoremstyle{thmstylethree}%

\begin{document}

\title[Calculation of tetraneutron-induced reaction cross sections
with optical and Hauser--Feshbach statistical models]{Calculation of tetraneutron-induced reaction cross sections
with optical and Hauser--Feshbach statistical models}


\author*[1]{\fnm{Hiroyuki} \sur{Fujioka}}\email{fujioka@phys.sci.isct.ac.jp}

\author[2]{\fnm{Toshihiko} \sur{Kawano}}\email{kawano@lanl.gov}

\affil*[1]{\orgdiv{Department of Physics}, \orgname{Institute of Science Tokyo}, \orgaddress{\city{Meguro}, \state{Tokyo}, \postcode{152-8551}, \country{Japan}}}

\affil[2]{\orgdiv{Theoretical Division}, \orgname{Los Alamos National Laboratory}, \orgaddress{\city{Los Alamos}, \state{NM}, \postcode{87545}, \country{USA}}}


\abstract{Interactions of tetraneutrons, which are assumed to be produced in the
nuclear fission process, with nuclei are studied in the framework of
optical and Hauser--Feshbach statistical models. It predicts a large
probability of $^{89}$Sr production for the tetraneutron-induced
reaction on $^{88}$Sr compared to other isotopes. The same technique
is applied to the tetraneutron-induced reaction on $^{27}$Al and the
hexaneutron-induced reaction on natural zinc to revisit two historical
multi-neutron experiments performed in the past.}

%\keywords{keyword1, Keyword2, Keyword3, Keyword4}



\maketitle

\section{Introduction}
\label{sec:Introduction}
The existence of a tetraneutron comprising four neutrons has been an
intriguing question for many decades~\cite{Marques:2021mqf}.
Recently, a peak structure was
observed in the $^{10}\mathrm{C}$ energy spectrum for the
$^7\mathrm{Li}({}^7\mathrm{Li},{}^{10}\mathrm{C})4\mathrm{n}$ reaction, leading
to an interpretation that a bound tetraneutron state with an energy of
$-0.42\pm 0.16\,\mathrm{MeV}$ relative to the four-neutron threshold
was produced~\cite{Faestermann:2022meh}.  It should be noted that
observations of a resonant state in different reactions were also
reported~\cite{Kisamori:2016jie,Duer:2022ehf},  
whereas they are challenged by several theoretical studies claiming the nonresonant nature of the four-neutron system~\cite{Deltuva2018,Higgins2020,Higgins2021,Lazauskas2023}.

Motivated by the finding, one of the authors (H.F.) and his
colleagues searched for a tetraneutron bound state, possibly emitted
in ternary fission of $^{235}\mathrm{U}$ induced by thermal
neutrons~\cite{Fujioka:2023gbr}. 
Recently Abdurrahman, et al.~\cite{Abdurrahman2024} 
reported in their fully microscopic study of fission dynamics that 
relatively high density of scission neutrons could be possible.
Different from the aforementioned
method to use nuclear reactions to populate tetraneutrons, the
existence of a tetraneutron can be confirmed if a reaction product in
a tetraneutron induced reaction with nuclei is identified.  This
method, named an activation probe, has been adopted mostly in the
early stage of searches for tetraneutrons as products in nuclear
fission or spallation~\cite{Marques:2021mqf}.  In
Ref~\cite{Fujioka:2023gbr}, they determined the upper limit of
$^{91}\mathrm{Sr}$ yield, which might be produced in the
$^{88}\mathrm{Sr}({}^4\mathrm{n}, \mathrm{n})$ reaction, inside the irradiated
$\mathrm{{}^{88}SrCO_3}$ sample, and, in turn, estimated the upper
limit of the emission rate of tetraneutron in $^{235}\mathrm{U}$
fission being $8\times 10^{-7}$.  However, there has been no
theoretical studies on tetraneutron-induced reaction such as
$({}^4\mathrm{n}, \mathrm{n})$, and the derived emission rate of tetraneutrons
is based on unsupported assumptions.  Indeed, the cross section of
$^{88}\mathrm{Sr}({}^4\mathrm{n}, \mathrm{n})$ was na\"{i}vely assumed to be
$50\,\mathrm{mb}$, which is taken from that of $(\alpha,\mathrm{n})$ reactions
on light nuclei, as done in Ref.~\cite{Schiffer1963}.

However, our current
understanding of two- and three-body nuclear force does not \textbf{allow} the 
existence of a physically observable resonant 
state~\cite{Lazauskas:2005ig,Hiyama:2016nwn,Higgins2021,Grigorenko2004,Timofeyuk2003,Pieper2003,Bertulani2003}, much less a bound state.
An approach alternative to experimental proof of nonexistence of such a bound state, 
which is virtually impossible, is accumulation of null results in search 
experiments with sufficiently high sensitivities.
From this viewpoint, the unsupported assumptions made in Ref.~\cite{Schiffer1963,Fujioka:2023gbr} should be carefully reexamined to draw a firm conclusion by providing reasonable estimates of tetraneutron interaction probabilities,
from which we can argue quantitatively non-observation of tetraneutrons.

In this paper we study the interaction between the hypothetical bound tetraneutron and
nuclei to estimate the production cross section of residual nuclei,
which might be observed in the tetraneutron experiments if they
exist. This is particularly important for estimating the upper limit
of tetraneutron emission rate in fission. Because there is no Coulomb
barrier, we expect the absorption cross section of tetraneutron could
be large at low energies, similar to neutron-induced reactions. Because
of the mass change ${}^AZ \to {}^{A+4}Z$, together with the expected
very weak binding of tetraneutron, the tetraneutron-induced reaction
may form a highly excited compound nucleus, and it de-excites by
sequential emission of particles such as neutron, proton,
$\alpha$-particle, and $\gamma$-ray. In this scenario, we are able to
apply the optical model for the tetraneutron absorption, and the
Hauser--Feshbach statistical theory~\cite{Hauser1952} for the compound
nucleus decay. First we demonstrate the tetraneutron-induced reaction
on $^{88}$Sr as in Ref~\cite{Fujioka:2023gbr}. Then we revisit two
reported experimental results by applying the same methodology; the
$^{28}$Mg production in a reactor by Schiffer and
Vandenbosch~\cite{Schiffer1963}, and the $^{72}$Zn production by the
24-GeV proton induced reaction on tungsten by
Detraz~\cite{Detraz1977}. Since Detraz's experiment claims a
possibility of existence of hexaneutron $^6$n, we extend our model to
the hexaneutron-induced reactions.

\section{Tetraneutron interaction with nucleus}
\label{sec:Calculation}
\subsection{Modeling framework}

Our assumed scenario is that first tetraneutrons ($^4$n) are produced
in the ternary fission process, then they interact with a target
nucleus nearby.  Naturally the produced $^4$n cannot have a high
kinetic energy.  Because $^4$n has no-charge, the absorption cross
section of $^4$n by the target nucleus might be large at low energies.
The reaction $^4{\rm n}+{}^AZ$ produces a compound nucleus
${}^{A+4}Z$, which could have a relatively high excitation energy of
20--30~MeV depending on the target nucleus and the $^4$n binding
energy.  Because of the expected high level densities of the compound
states, we apply the optical model and the Hauser--Feshbach statistical
model~\cite{Hauser1952} for the $^4$n induced reaction on target
nuclei.

We make a few plausible assumptions that might be uncommon for nucleon
or composite particle induced nuclear reactions. We do not consider
direct reactions, such as the sequential stripping of neutrons
consisting of $^4$n by the target. This is a multi-step process, and
its amplitude should be very small. We also do not include the
pre-equilibrium process because of the energy region of our
interest.
Although these reaction mechanisms may play some roles in estimating the activation cross sections when the $^4$n kinetic energy is rather high (e.g. more than 10~MeV), such the scenario is hard to envisage in our regime, because a typical average energy of a single neutron in fission is only 2~MeV.
Finally, we ignore $^4$n emission from the compound
nucleus. In other words, the compound elastic scattering does not take
place. This ignores the time-reversal property of the nuclear
reaction. However, because the excitation energy of the compound
nucleus is so high and there are an enormous number of decay channels,
the compound elastic probability becomes extremely small. 
The probability of compound nucleus decaying into the incoming channel is negligible due to the high level density in the residual nucleus of single neutron emission. 
For the same
reason, we do not include the width fluctuation correction~\cite{Moldauer1961,Moldauer1964,Kawano2015},
because the negligible cross section of compound elastic scattering does not change any reaction channels.
Based on
these assumptions, the formation of compound nucleus and the decay
process are decoupled, except for the angular momentum conservation.

\subsection{Absorption of tetraneutron}

Since the nuclear interaction between $^4$n and a target nucleus is
unknown, we take a look into a similar compound nuclear reaction with
a composite particle such as deuteron, $\alpha$-particle, and so
on. Our naive assumption is that $^4$n is absorbed by a target nucleus
and its energy spreads immediately to achieve an equilibrium
condition. We employ the optical model for the formation of the
compound nucleus.

We estimate the optical potential for $^4$n by a simple folding
approach by Watanabe~\cite{Watanabe1958} and
Madland~\cite{Madland1988}. The real potential depth at the incident
energy $E$ is given by
\begin{equation}
  V_{{}^4{\rm n}}(E) \simeq 4 V_{\rm n}(E/4) \ ,
  \label{eq:4nfoldingV}
\end{equation} 
where $V_{\rm n}$ is the real part of optical potential for a single
neutron. The Woods-Saxon form is adopted for the radial dependence,
and the geometrical parameters (radius and diffuseness) are the same
as those for the neutron optical potential; 
the radius parameters
for the real volume, imaginary volume, and imaginary surface potentials 
are $r_v=1.213$~fm, $r_w=1.213$~fm, and $r_s=1.272$~fm. The diffuseness
parameters for these potential components are 
$a_v=0.665$~fm, $a_w=0.665$~fm, and $a_s=0.530$~fm.
The imaginary surface term
$W_S$ and the volume term $W_V$ are calculated in the same way,
\begin{eqnarray}
  W_{S,{}^4{\rm n}}(E) &\simeq& 4 W_{S,\rm n}(E/4) \ , \\
  W_{V,{}^4{\rm n}}(E) &\simeq& 4 W_{V,\rm n}(E/4) \ ,
 \label{eq:4nfoldingW}
\end{eqnarray} 
while we omit the spin-orbit potential as $^4$n has zero intrinsic
spin. We adopt the Koning--Delaroche global potential~\cite{Koning2003}
for the neutron-target optical potential.

Typical neutron optical potentials have a real depth of about 50~MeV
or so.  Watanabe--Madland's prescription gives the $^4$n potential
depth to be 200~MeV, which tends to be deeper than the optical
potentials for the $\alpha$-particle.  For example, in the
$\alpha+{}^{88}$Sr reaction case, the typical values are
171~MeV~\cite{Avrigeanu2014}, 153~MeV~\cite{Nolte1987}, and
193~MeV~\cite{Lemos1972}. However, the calculated $\alpha$-particle
absorption cross section with the Watanabe--Madland method is not so
different from the reported $\alpha$-particle potentials.  For the
10-MeV $\alpha$-particle induced reaction on $^{88}$Sr, the calculated
absorption cross sections by the aforementioned three
potentials~\cite{Avrigeanu2014,Nolte1987,Lemos1972} are, respectively,
12.4, 36.8, and 15.2~mb, while the Watanabe--Madland method with the
Koning--Delaroche potential gives 21.5~mb.

Cross sections for the compound nucleus formation are calculated by
the $^4$n transmission coefficients $T_l$, where $l$ is the orbital
angular momentum coupled with the target spin $I_t$ to form the spin of
initial compound state, ${\bf J} = {\bf l} + {\bf I}_t$ with an
appropriate parity $\Pi$. The partial cross section for populating 
the $J\Pi$ compound state is given by
\begin{equation}
 \sigma_{J\Pi} = \frac{2J+1}{2I_t+1} \frac{\pi}{k^2_{{}^4{\rm n}}} \sum_l T_l f(\Pi;\Pi_t,l) \ ,
 \label{eq:sigma}
\end{equation}
where $k_{{}^4{\rm n}}$ is the wave-number of $^4$n, $\Pi_t$ is the 
target parity, and
\begin{equation}
 f(\Pi;\Pi_t,l) =
   \left\{
    \begin{array}{ll}
       1 & (\Pi = \Pi_t (-)^l) \\
       0 & (\mbox{otherwise})
    \end{array}
   \right.
\end{equation}
ensures the parity conservation in the summation. The total compound
nucleus formation cross section is given by $\sigma_{\rm CN} =
\sum_{J\Pi} \sigma_{J\Pi}$.

\subsection{Statistical decay of compound nucleus}

The formed compound state for a given $J\Pi$ decays by multiple
emission of the neutron, proton, $\alpha$-particle, $\gamma$-ray, and
so on, which are characterized by the particle ($\gamma$-ray)
transmission coefficients, the level densities, and the nuclear
structure of low-lying discrete levels.

The optical and Hauser--Feshbach statistical models are performed with
the CoH$_3$ code~\cite{Kawano2019}, which is specifically modified to
handle the $^4$n channel. The optical potentials for the neutron and
proton decay channels are taken from
Koning and Delaroche~\cite{Koning2003}. For the $\alpha$-particle, we
employ the global potentials of Avrigeanu {\it et
  al.}~\cite{Avrigeanu2014}. The decay channels for deuteron, triton,
and $^3$He are ignored for the reactions on medium to heavy targets,
since the production cross sections for these particles are expected
to be negligible. In the lighter target cases, these channels are
explicitly included. The deuteron optical potential is taken from
Bojowald {\it et al.}~\cite{Bojowald1988}.  The optical potentials for
the triton and $^3$He channels are taken from Becchetti and
Greenlees~\cite{Becchetti1969b}. The $\gamma$-ray transmission
coefficient is calculated by the giant dipole resonance model, and the
$\gamma$-ray strength function is taken from Kopecky and
Uhl~\cite{Kopecky1990}. We calculate the level density of the residual
nuclei with the Gilbert--Cameron composite formula~\cite{Gilbert1965,
  Kawano2006}, and take the known discrete level data from
RIPL-3~\cite{RIPL3}. Besides the incoming channel, the modeling
framework is pretty standard and this well-established framework
has been applied to many nuclear reactions in the past.

Uncertainty in the binding energy of $^4$n causes undetermined total
excitation energy of the compound nucleus. For the $^4{\rm n} + {}^AZ$
reaction, the excitation energy of ${}^{A+4}Z$ is
\begin{equation}
  E_x = E + \Delta M({}^AZ) + \Delta M({}^4{\rm n}) - \Delta M({}^{A+4}Z) \ ,
\end{equation}
where $E$ is the incident energy in the center-of-mass and $\Delta
M({}^AZ)$ is the mass excess. Because the $^4$n mass excess
$\Delta({}^4{\rm n})$ is unknown, we perform the Hauser--Feshbach
calculation by assuming several values of the binding energy varying
from 100~keV to 5~MeV, which gives $\Delta({}^4{\rm n}) = 4\Delta(n) -
B({}^4{\rm n}) = 32.3 - (0.1\sim 5)$~MeV.

\section{Results and Discussion}
\label{sec:Results}
\subsection{Tetraneutron-induced reaction on strontium-88}

\begin{figure}
  \begin{center}
    \includegraphics[width=\columnwidth]{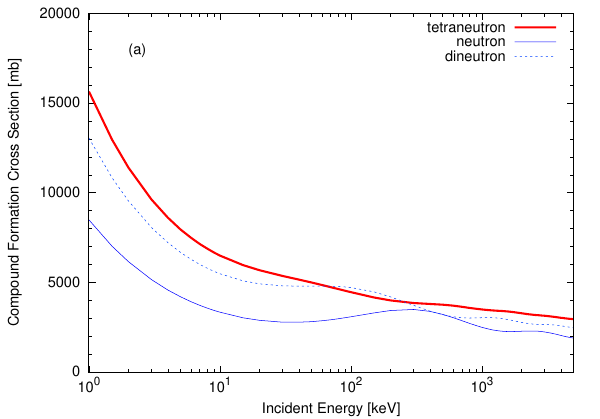}\\
    \includegraphics[width=\columnwidth]{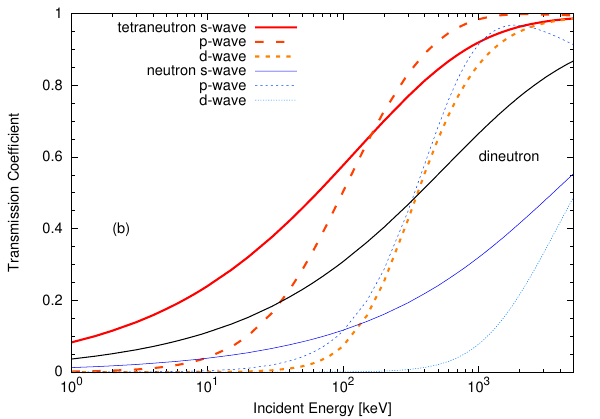}
  \end{center}
  \caption{Upper panel (a): calculated compound formation cross sections for
    the $^4$n, $^2$n, and neutron induced reactions on $^{88}$Sr.
    Lower panel (b): transmission coefficients for each of partial waves.
    Only $s$-wave is shown for the $^2$n case.}
  \label{fig:reactionSr88}
\end{figure}

First we performed the optical model calculation for the $^4$n
scattering from $^{88}$Sr. Because the kinetic energy of $^4$n
produced by the ternary fission is also unknown, we assume it would be
somewhere in the keV to MeV region. The kinetic energy could have a
distribution too. The top panel of Fig.~\ref{fig:reactionSr88} shows
the calculated compound nucleus formation cross section for $^{88}$Sr
at low energies by employing
Eqs.~(\ref{eq:4nfoldingV})--(\ref{eq:4nfoldingW}).  We also depicted
the neutron-induced reaction case for a comparison. As the energy
approaches zero, both curves increase rapidly due to the kinematic
factor of $\pi/k^2$ in Eq.~(\ref{eq:sigma}).  The wave-number of
$^4$n, $k_{{}^4{\rm n}}$, is two times larger than the neutron
wave-number $k_{\rm n}$ at the same incident energy, because of the
difference in the reduced mass of relative motion. This implies $^4$n
has less probability of forming a compound nucleus, while the
calculated result is opposite. The reason is the $^4$n transmission
coefficient plotted in the lower panel. Each of the partial waves for
$^4$n rises faster than the neutron transmission coefficients, which
results in the larger compound formation cross section.  Note that in
neutron case, the transmission coefficients for the two spin states,
$T_{l+1/2}$ and $T_{l-1/2}$, are shown by the weighted averaged
\begin{equation}
  T_l = \frac{(l+1)T_{l+1/2} + lT_{l-1/2}}{2l+1} 
\end{equation}
for $l \ge 1$.

In the energy range 1~keV to 1~MeV, a few partial waves already
constitute the total reaction cross section.  We don't know exactly
what is an average energy of $^4$n that is interacting with the target
nucleus. When $\sim 1$~keV, the compound nucleus spin will be
identical to the target nucleus spin, as the $s$-wave is dominant. At
higher energies, the populated spin states form a distribution by the
particle transmission coefficient for each of partial waves. This is
particularly important to calculate the statistical decay of produced
compound nucleus, which could have a very high excitation energy
compared to a single low-energy neutron incident case. In the case of
de-excitation from a highly excited state with a very narrow spin
distribution, the spin-parity conservation restricts the allowed final
states tighter than nucleon-induced reaction cases.

In a case the nuclear fission produces dineutrons ($^2$n), they also contribute to the compound nucleus formation. We calculated the dineutron absorption cross section and the transmission coefficient in the same manner as $^4$n, and shown in Fig.~\ref{fig:reactionSr88}. The assumed $^2$n binding energy is zero. Although only the $s$-wave transmission coefficient is shown for the sake of better visibility, they lie between the single n and $^4$n cases.

\begin{figure}
  \begin{center}
    \includegraphics[width=\columnwidth]{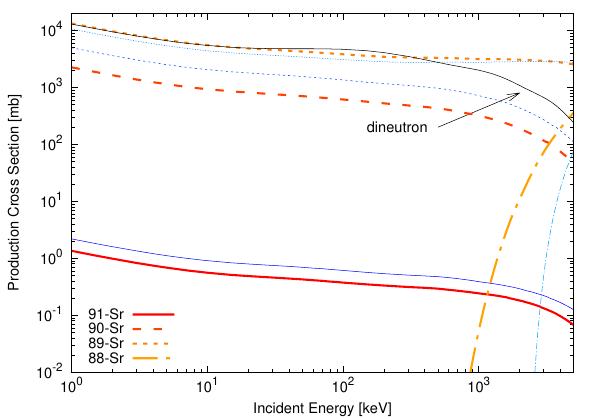}
  \end{center}
  \caption{Calculated production cross sections for different
           Sr isotopes. Two binding energies are assumed;
           0.42~MeV (thick curves) and 2~MeV (thin curves).
           The production cross section
            for $^{88}$Sr($^2$n,n)$^{89}$Sr is indicated by the arrow.}
  \label{fig:production1}
\end{figure}

An absorbed $^4$n by $^{88}$Sr forms $^{92}$Sr, and it decays by
emitting several neutrons and $\gamma$-rays until it exhausts the
initial excitation energy.  Figure~\ref{fig:production1} shows the
production cross sections of $^{91}$Sr, $^{90}$Sr, $^{89}$Sr, and
$^{88}$Sr as a function of incident energy, which are for the 1n, 2n,
3n, and 4n emission channels. The thick curves are for the $B({}^4{\rm
  n})=0.42$~MeV case~\cite{Faestermann:2022meh}, and the thin curves are for 2~MeV. 
  The 2~MeV was chosen to show how these production cross sections are sensitive to $B({}^4{\rm n})$, since the shape of each of the production cross sections is similar, but it changes the magnitude.
The production
of $^{92}$Sr is extremely small, and not shown in this figure. For the
both binding energy cases, the most probable reaction in this energy
range is the 3-neutron emission channel, and the 2-neutron
emission channel follows. This implies the production of $^{89}$Sr by a
fission-produced $^4$n is several times larger than $^{90}$Sr
regardless of the assumed binding energy.

In particular, the cross section of $\mathrm{{}^{88}Sr({}^4n,n){}^{91}Sr}$ is found to be 1--2 orders of magnitude smaller than that assumed in Ref.~\cite{Fujioka:2023gbr}.
This implies that the reported upper limit of the tetraneutron emission rate in $\mathrm{^{235}U}$ fission could have been underestimated by the same amount.

\begin{figure}
  \begin{center}
    \includegraphics[width=\columnwidth]{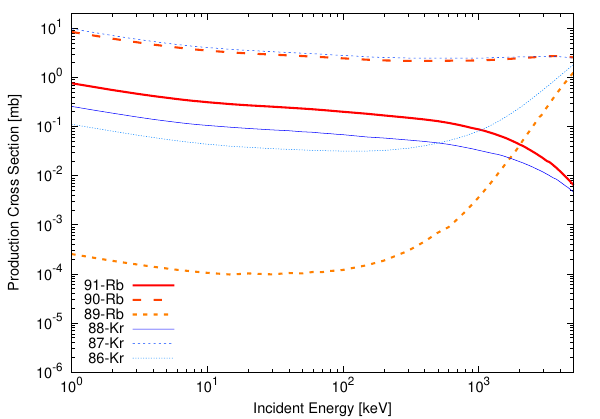}
  \end{center}
  \caption{Calculated production cross sections for 
           charged-particle production channels. 
           The assumed binding energy is 0.42~MeV.}
  \label{fig:production2}
\end{figure}

Production cross sections for charged particle emission channels are
much smaller than the neutron-only cases, so we show them separately
in Fig.~\ref{fig:production2}. They are for the ($^4$n,$x$np) and
($^4$n,$x$n$\alpha$) reactions, where $x=0,1$, and 2. As demonstrated
in Fig.~\ref{fig:production1}, change in the binding shifts the entire
functions but the shape remains similar.  We show the $B({}^4{\rm
  n})=0.42$~MeV case only. The np and n$\alpha$ emission channels have
the largest cross sections, albeit they are at most 10~mb or so.

Figure~\ref{fig:production1} also includes the $^{89}$Sr production cross section {\it via} ($^2$n,n) reaction, which turned out to be in the same magnitude as the ($^4$n,3n) cross section. This may make the detection of $^4$n more challenging. One possible way to distinguish these reaction channels is to measure the decay of $^{90}$Sr, since the $^2$n capture cross section that makes $^{90}$Sr is a few orders of magnitude smaller than the ($^4$n,2n) reaction. 

It is not so straightforward if a resonance state is formed by an unbound $^4$n within our model. Although it is a rather crude estimate, we would assume a compound nucleus is produced by simultaneous absorption of four neutrons, and that state decays in a similar manner as the statistical theory. Since the total spin of four neutrons is unknown, the total spin of the compound nucleus could be different from the tetraneutron case. Besides the spin, the initial excitation energy is the same as the case where the tetraneutron binding energy is zero. The production of $^{89}$Sr might be slightly higher than the $B({}^4{\rm n})=0.42$~MeV case.

\begin{figure}
  \begin{center}
    \includegraphics[width=\columnwidth]{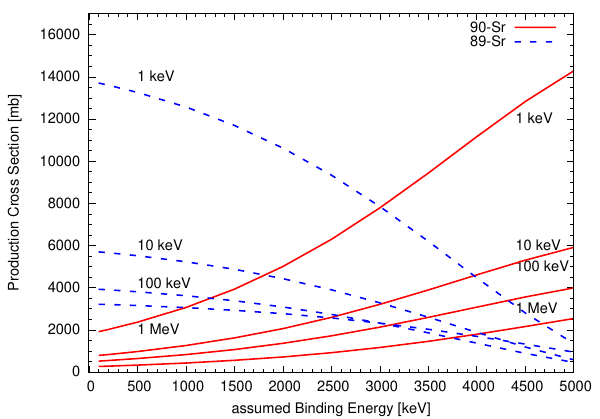}
  \end{center}
  \caption{Production cross section of $^{90}$Sr and $^{89}$Sr as a
           function of assumed binding energy. The different curves 
           are for the incident energies of 1, 10, 100, and 1000~keV.}
  \label{fig:binding}
\end{figure}

As shown in Fig.~\ref{fig:production1}, the 2n and 3n emissions are
the most likely process after $^4$n is absorbed by $^{88}$Sr.  The
ratio of production of $^{89}$Sr and $^{90}$Sr depends on the assumed
binding energy. In Fig.~\ref{fig:binding} we plot the 2n and 3n
emission cross sections as a function of the binding energy assumed.
Since the ratio also depends on the incident energy, we depicted the
results at 1, 10, 100, and 1000~keV incident energies.

\subsection{Tetraneutron-induced reaction on aluminum}

\begin{figure}
  \begin{center}
    \includegraphics[width=\columnwidth]{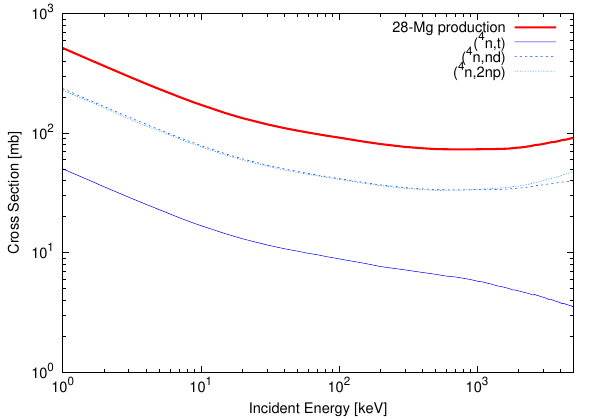}
  \end{center}
  \caption{Calculated cross sections 
           of ($^4$n,t), ($^4$n,nd), and ($^4$n,2np) reactions on $^{27}$Al,
           and the total $^{28}$Mg production cross section.
           The assumed binding energy is 0.42~MeV.}
  \label{fig:production3}
\end{figure}

We revisit the historical experiment by Schiffer and
Vandenbosch~\cite{Schiffer1963}.  They looked for a possibility of
$^{28}$Mg production by the ${}^4{\rm n} + {}^{27}{\rm Al}$ reaction
in a research reactor. Our computational procedure is the same as the
$^{88}$Sr case, while we included the deuteron, triton, and $^3$He
emission channels because ($^4$n,t), ($^4$n,nd), and ($^4$n,2np)
produce the same residual nucleus of $^{28}$Mg.  The calculated
$^{28}$Mg production cross sections are shown in
Fig.~\ref{fig:production3}, 
where the assumed binding energy is 0.42~MeV.
Although the ($^4$n,t) cross section
tends to be lower than the other reactions, these three reaction
channels contribute comparably to the total $^{28}$Mg production. It
seems the production cross section is reasonably large to detect it
experimentally.  Schiffer and Vandenbosch assumed the $^{28}$Mg
production cross section is 40~mb regardless of the $^4$n kinetic
energy. Our prediction is one order magnitude higher than their
assumption, which implies that the $^4$n production per fission could
be one order magnitude smaller than their estimated value of $5\times
10^{-9}$.

 It would be interesting to repeat the experiment of Schiffer and Vandenbosch with more advanced activation measurement techniques, as the $^{28}$Mg production is another candidate to confirm the existence of $^4$n. The estimated $^{28}$Mg production cross section is not so sensitive to the assumed binding energy, because the total excitation energy is already 30~MeV or so.

\subsection{Hexaneutron-induced reaction on zinc}

In the same framework, it is possible to calculate the statistical
decay of compound nucleus produced by a hexaneutron, $^6$n. We
calculate the $^6$n induced reaction on natural Zn to revisit the
measurement by Detraz~\cite{Detraz1977}. Because the $^6$n binding
energy, if it is a bound system, is unknown, we set it to be
1~MeV. This energy is rather insensitive to our results as far as the
actual binding is not significantly stronger, because the excitation
energy of $^{A+6}$Zn is much higher. For example, the excitation
energy for the reaction $^6{\rm n}+{}^{64}{\rm Zn}$ is 51~MeV, and
$^6{\rm n}+{}^{70}{\rm Zn}$ is 40~MeV.

The effective kinetic energy of the interaction in Detraz's
measurement is also unknown. We first calculate the reaction for the
1-keV incident case, which could be way lower than an expected energy
in the reaction of 24-GeV proton impinged on tungsten.
Although the choice of lower limit of 1~keV is rather arbitrary, an existence of the hexaneutron could be already reported if the actual kinetic energy is lower, since the interaction cross section obeys the $1/v$-law.
When the kinetic energy increases, more neutrons are evaporated from the formed
compound nucleus, hence lighter Zn isotopes are preferably produced.
The choice of 1~keV is to illustrate the heavy-side limit of produced
Zn isotopes.

\begin{figure}
  \begin{center}
    \includegraphics[width=\columnwidth]{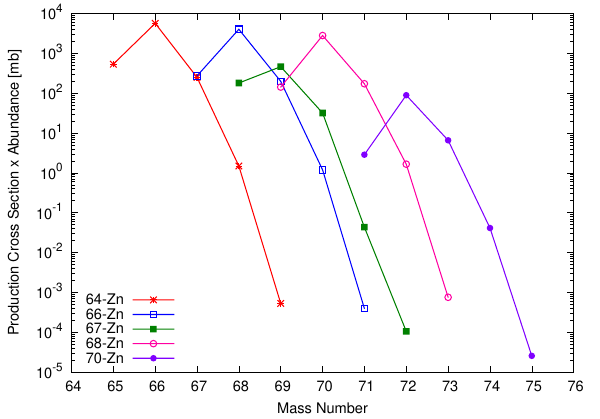}
  \end{center}
  \caption{Production cross sections of Zn isotopes for the $^6$n
           reaction on the natural Zn target. The plotted values include
           the natural abundance of each Zn isotope.
           An assumed binding energy is 1~MeV.}
  \label{fig:hexa}
\end{figure}

The calculated production cross sections for each of Zn isotopes are
shown in Fig.~\ref{fig:hexa}, which includes the natural
abundance. The production of $^{72}$Zn, which Detraz claims that it
was produced by $^6$n, is mainly from $^{70}$Zn, and a small fraction
by $^{68}$Zn. When the kinetic energy is larger, all the curves in
Fig.~\ref{fig:hexa} move to the left hand side, and the production of
$^{72}$Zn becomes less probable. For example, the calculated
production cross section for the $^{70}$Zn target is 23~mb at 100~keV,
19~mb at 1~MeV, and 5.4~mb at 10~MeV. It may not be so incorrect to
assume that the kinetic energy of $^6$n, if exists, is quite large for
the 24-GeV proton measurement, which results in a much smaller
production cross section of $^{72}$Zn. Having said that, we do not
argue that detection is impossible. It just implies that the
production of $^{72}$Zn by $^6$n is unlikely.

\section{Conclusion}
\label{sec:Conclusion}
We calculated the tetraneutron ($^4$n) induced reaction cross sections
on several targets by employing the optical and Hauser--Feshbach
statistical models. This aims at providing fundamental information for
search of $^4$n when expected residual products of $^4$n induced
reaction are experimentally looked for, albeit some uncertain
quantities such as the $^4$n binding energy are involved in the
developed model. The optical potential for $^4$n was constructed by
the simple folding approach proposed by Watanabe~\cite{Watanabe1958}
and Madland~\cite{Madland1988}. The calculated $^4$n absorption cross
section for the $^{88}$Sr target was 5 to 15~b in the keV region,
where we assumed that $^4$n's are born in the ternary fission
because the neutron density at scission might be 
high enough as implied by the microscopic fission model~\cite{Abdurrahman2024}.
The formed $^{92}$Sr compound nucleus decays mainly by emitting several
neutrons, and the strongest reaction was found to be ($^4$n,3n) that
produces $^{89}$Sr.  Although we included the proton and
$\alpha$-particle emissions, their cross sections were much smaller
than the neutron emission cases.

The same methodology was applied to other reactions to revisit two
historical experiments. First, we examined Schiffer and Vandenbosch's
experiment~\cite{Schiffer1963}. They assumed that the interaction of
$^4$n with $^{27}$Al produces $^{28}$Mg in a reactor.  Our calculated
$^{28}$Mg production cross section was much larger than their assumed
value of 40~mb, which implies their estimated $^4$n production rate
could be lower. Second, we extended our calculation framework to
the hexaneutron-induced reaction, and examined the measurement of
Detraz~\cite{Detraz1977}. Although he claimed that $^{72}$Zn was
produced by $^6$n, our results suggest $^{72}$Zn production by $^6$n
is unlikely.

Finally, we emphasize that we are not claiming
the existence of $^4$n in this study, but our aim is to provide
crucial information of $^4$n nuclear reaction rate to determine the
lower limit in the measurement when the activation probe technique is
performed. Although our estimates may still have some uncertainties
due to the assumed binding energy, the endeavor to experimentally
determine the probability of existence is impossible without the
estimation of reaction rates.

\bmhead{Acknowledgements}

TK carried out this work under the auspices of the National Nuclear
Security Administration of the U.S. Department of Energy at Los Alamos
National Laboratory under Contract No. 89233218CNA000001.
HF acknowledges the financial support by the Sumitomo Foundation (grant number 2402570).
\bmhead{Data Availability Statement}

All the data produced in this study are already shown in the figures.
We will make all the numerical data available upon researchers'
request.

\bmhead{Code Availability Statement}

All the calculations were made with the statistical Hauser--Feshbach
code CoH3, which was released as open-source, and publicly available
at \url{https://github.com/toshihikokawano/coh3}
\bibliography{ref}

@article{Marques:2021mqf,
    author = "Marqu\'{e}s, F. Miguel and Carbonell, Jaume",
    title = "{The quest for light multineutron systems}",
    doi = "10.1140/epja/s10050-021-00417-8",
    journal = "Eur. Phys. J. A",
    volume = "57",
    number = "3",
    pages = "105",
    year = "2021"
}

@article{Lazauskas:2005ig,
	author = "Lazauskas, Rimantas and Carbonell, Jaume",
	title = "{Is a physically observable tetraneutron resonance compatible with realistic nuclear interactions?}",
	doi = "10.1103/PhysRevC.72.034003",
	journal = "Phys. Rev. C",
	volume = "72",
	pages = "034003",
	year = "2005"
}

@article{Hiyama:2016nwn,
    author = "Hiyama, E. and Lazauskas, R. and Carbonell, J. and Kamimura, M.",
    title = "{Possibility of generating a 4-neutron resonance with a $T=3/2$ isospin 3-neutron force}",
    eprint = "1604.04363",
    archivePrefix = "arXiv",
    primaryClass = "nucl-th",
    doi = "10.1103/PhysRevC.93.044004",
    journal = "Phys. Rev. C",
    volume = "93",
    number = "4",
    pages = "044004",
    year = "2016"
}

@article{Faestermann:2022meh,
	author = {Faestermann, Thomas and Bergmaier, Andreas and Gernh\"auser, Roman and Koll, Dominik and Mahgoub, Mahmoud},
	title = "{Indications for a bound tetraneutron}",
	doi = "10.1016/j.physletb.2021.136799",
	journal = "Phys. Lett. B",
	volume = "824",
	pages = "136799",
	year = "2022"
}

@article{Kisamori:2016jie,
  title = {Candidate Resonant Tetraneutron State Populated by the $^{4}\mathrm{He}(^{8}\mathrm{He},^{8}\mathrm{Be})$ Reaction},
  author = {Kisamori, K. and Shimoura, S. and Miya, H. and Michimasa, S. and Ota, S. and Assie, M. and Baba, H. and Baba, T. and Beaumel, D. and Dozono, M. and Fujii, T. and Fukuda, N. and Go, S. and Hammache, F. and Ideguchi, E. and Inabe, N. and Itoh, M. and Kameda, D. and Kawase, S. and Kawabata, T. and Kobayashi, M. and Kondo, Y. and Kubo, T. and Kubota, Y. and Kurata-Nishimura, M. and Lee, C. S. and Maeda, Y. and Matsubara, H. and Miki, K. and Nishi, T. and Noji, S. and Sakaguchi, S. and Sakai, H. and Sasamoto, Y. and Sasano, M. and Sato, H. and Shimizu, Y. and Stolz, A. and Suzuki, H. and Takaki, M. and Takeda, H. and Takeuchi, S. and Tamii, A. and Tang, L. and Tokieda, H. and Tsumura, M. and Uesaka, T. and Yako, K. and Yanagisawa, Y. and Yokoyama, R. and Yoshida, K.},
  journal = {Phys. Rev. Lett.},
  volume = {116},
  issue = {5},
  pages = {052501},
  numpages = {5},
  year = {2016},
  month = {Feb},
  publisher = {American Physical Society},
  doi = {10.1103/PhysRevLett.116.052501},
  url = {https://link.aps.org/doi/10.1103/PhysRevLett.116.052501}
}

@article{Duer:2022ehf,
  author = {Duer, M. and Aumann, T. and Gernh\"{a}user, R. and Panin, V. and Paschalis, S. and Rossi, D. M. and Achouri, N. L. and Ahn, D. and Baba, H. and Bertulani, C. A. and B\"{o}hmer, M. and Boretzky, K. and Caesar, C. and Chiga, N. and Corsi, A. and Cortina-Gil, D. and Douma, C. A. and Dufter, F. and Elekes, Z. and Feng, J. and Fern\'{a}ndez-Dom\'{\i}nguez, B. and Forsberg, U. and Fukuda, N. and Gasparic, I. and Ge, Z. and Gheller, J. M. and Gibelin, J. and Gillibert, A. and Hahn, K. I. and Hal\'{a}sz, Z. and Harakeh, M. N. and Hirayama, A. and Holl, M. and Inabe, N. and Isobe, T. and Kahlbow, J. and Kalantar-Nayestanaki, N. and Kim, D. and Kim, S. and Kobayashi, T. and Kondo, Y. and K\"{o}rper, D. and Koseoglou, P. and Kubota, Y. and Kuti, I. and Li, P. J. and Lehr, C. and Lindberg, S. and Liu, Y. and Marqu\'{e}s, F. M. and Masuoka, S. and Matsumoto, M. and Mayer, J. and Miki, K. and Monteagudo, B. and Nakamura, T. and Nilsson, T. and Obertelli, A. and Orr, N. A. and Otsu, H. and Park, S. Y. and Parlog, M. and Potlog, P. M. and Reichert, S. and Revel, A. and Saito, A. T. and Sasano, M. and Scheit, H. and Schindler, F. and Shimoura, S. and Simon, H. and Stuhl, L. and Suzuki, H. and Symochko, D. and Takeda, H. and Tanaka, J. and Togano, Y. and Tomai, T. and T\"{o}rnqvist, H. T. and Tscheuschner, J. and Uesaka, T. and Wagner, V. and Yamada, H. and Yang, B. and Yang, L. and Yang, Z. H. and Yasuda, M. and Yoneda, K. and Zanetti, L. and Zenihiro, J. and Zhukov, M. V.},
  title = {Observation of a correlated free four-neutron system},
  doi = {10.1038/s41586-022-04827-6},
  journal = {Nature},
  volume = {606},
  number = {7915},
  pages = {678--682},
  year = {2022}
}

@article{Fujioka:2023gbr,
    author = "Fujioka, Hiroyuki and Tomomatsu, Ryutaro and Takamiya, Koichi",
    title = "{Search for particle-stable tetraneutrons in thermal fission of $^{235}$U}",
    doi = "10.1103/PhysRevC.108.054004",
    journal = "Phys. Rev. C",
    volume = "108",
    number = "5",
    pages = "054004",
    year = "2023"
}

@article{Hauser1952,
  title = {The Inelastic Scattering of Neutrons},
  author = {Hauser, Walter and Feshbach, Herman},
  journal = {Phys. Rev.},
  volume = {87},
  issue = {2},
  pages = {366--373},
  numpages = {0},
  year = {1952},
  month = {Jul},
  publisher = {American Physical Society},
  doi = {10.1103/PhysRev.87.366},
  url = {http://link.aps.org/doi/10.1103/PhysRev.87.366}
}

@article{Moldauer1961,
  title = {Theory of Average Neutron Reaction Cross Sections in the Resonance Region},
  author = {Moldauer, P. A.},
  journal = {Phys. Rev.},
  volume = {123},
  issue = {3},
  pages = {968 -- 978},
  numpages = {0},
  year = {1961},
  month = {Aug},
  publisher = {American Physical Society},
  doi = {10.1103/PhysRev.123.968},
  url = {http://link.aps.org/doi/10.1103/PhysRev.123.968}
}

@article{Moldauer1964,
  title = {Statistical Theory of Nuclear Collision Cross Sections},
  author = {Moldauer, P. A.},
  journal = {Phys. Rev.},
  volume = {135},
  issue = {3B},
  pages = {B642 -- B659},
  numpages = {0},
  year = {1964},
  month = {Aug},
  publisher = {American Physical Society},
  doi = {10.1103/PhysRev.135.B642},
  url = {http://link.aps.org/doi/10.1103/PhysRev.135.B642}
}

@article{kawano2015,
  title = {Random-matrix approach to the statistical compound nuclear reaction at low energies using the {Monte Carlo} technique},
  author = {Kawano, T. and Talou, P. and Weidenm\"uller, H. A.},
  journal = {Phys. Rev. C},
  volume = {92},
  issue = {4},
  pages = {044617},
  numpages = {16},
  year = {2015},
  month = {Oct},
  publisher = {American Physical Society},
  doi = {10.1103/PhysRevC.92.044617},
  url = {http://link.aps.org/doi/10.1103/PhysRevC.92.044617}
}

@article{Watanabe1958,
  title = {High energy scattering of deuterons by complex nuclei},
  author = {Watanabe, Shigueo},
  journal = {Nuclear Physics},
  volume = {8},
  pages = {484 -- 492},
  year = {1958},
  issn = {0029-5582},
  doi = {10.1016/0029-5582(58)90180-9},
  url = {http://www.sciencedirect.com/science/article/pii/0029558258901809},
}

@article{Madland1988,
  title = {Recent results in the development of a global medium-energy nucleon-nucleus optical-model potential},
  author = {Madland, D.G.},
  pages = {103--116},
  year = {1988},
  journal = {Proc. Specialists' meeting on preequilibrium nuclear reactions},
  publisher = {OECD/NEA},
  volume = {NEANDC-245},
  note = {10 -- 12 Feb. 1988, Vienna, Austria},
}

@article{Koning2003,
  title = {{Local and global nucleon optical models from 1 keV to 200 MeV}},
  author = {Koning, A. J.  and Delaroche, J.-P.},
  journal = {Nuclear Physics A},
  volume = {713},
  number = {3},
  pages = {231 -- 310},
  year = {2003},
  issn = {0375-9474},
  doi = {10.1016/S0375-9474(02)01321-0},
  url = {http://www.sciencedirect.com/science/article/pii/S0375947402013210},
}

@article{Avrigeanu2014,
  title = {Further explorations of the $\ensuremath{\alpha}$-particle optical model potential at low energies for the mass range ${A}\ensuremath{\approx}45$--209},
  author = {Avrigeanu, V. and Avrigeanu, M. and M\u{a}n\u{a}ilescu, C.},
  journal = {Phys. Rev. C},
  volume = {90},
  issue = {4},
  pages = {044612},
  numpages = {13},
  year = {2014},
  month = {Oct},
  publisher = {American Physical Society},
  doi = {10.1103/PhysRevC.90.044612},
  url = {https://link.aps.org/doi/10.1103/PhysRevC.90.044612}
}

@article{Nolte1987,
  title = {Global optical potential for \ensuremath{\alpha} particles with energies above 80 {MeV}},
  author = {Nolte, M. and Machner, H. and Bojowald, J.},
  journal = {Phys. Rev. C},
  volume = {36},
  issue = {4},
  pages = {1312--1316},
  numpages = {0},
  year = {1987},
  month = {Oct},
  publisher = {American Physical Society},
  doi = {10.1103/PhysRevC.36.1312},
  url = {https://link.aps.org/doi/10.1103/PhysRevC.36.1312}
}

@techreport{Lemos1972,
  author = {Lemos, Orlando F.},
  title = {Diffusion elastique de particules alpha de 21 a 29.6 {MeV} sur des noyaux de la region {Ti-Zn}},
  institution = {Orsay report},
  year = {1972},
  number = {series A136}
}

@article{Kawano2019,
  title = {{CoH$_3$}: The Coupled-Channels and {Hauser--Feshbach} Code},
  author = {Kawano, Toshihiko},
  journal = {Springer Proceedings in Physics},
  year = {2021},
  volume = {254},
  pages = {27 -- 34},
  note = {{CNR2018}: International Workshop on Compound Nucleus and Related Topics, LBNL, Berkeley, CA, USA, September 24 -- 28, 2018, J. Escher, Y. Alhassid, L.A. Bernstein, D. Brown, C. Fr\"{o}hlich,  P. Talou, W. Younes (Eds.)},
  isbn = {978-3-030-58081-0},
  issn = {0930-8989},
  doi = {10.1007/978-3-030-58082-7_3},
  url = {https://doi.org/10.1007/978-3-030-58082-7_3}
}

@article{Kopecky1990,
  title = {Test of gamma-ray strength functions in nuclear reaction model calculations},
  author = {Kopecky, J. and Uhl, M.},
  journal = {Phys. Rev. C},
  volume = {41},
  issue = {5},
  pages = {1941 -- 1955},
  numpages = {0},
  year = {1990},
  month = {May},
  publisher = {American Physical Society},
  doi = {10.1103/PhysRevC.41.1941},
  url = {http://link.aps.org/doi/10.1103/PhysRevC.41.1941}
}

@article{Gilbert1965,
  title = {A composite nuclear-level density formula with shell corrections},
  author = {Gilbert, A. and Cameron, A. G. W.},
  journal = {Can. J. Phys.},
  volume = {43},
  pages = {1446 -- 1496},
  year = {1965},
  doi = {10.1139/p65-139},
  url = {http://www.nrcresearchpress.com/doi/abs/10.1139/p65-139\#.V6IU247raQ8}
}

@article{Kawano2006,
  title = {Phenomenological Nuclear Level Densities using the {KTUY05} Nuclear Mass Formula for Applications Off-Stability},
  author = {Kawano,  Toshihiko  and  Chiba, Satoshi and  Koura, Hiroyuki},
  journal = {Journal of Nuclear Science and Technology},
  volume = {43},
  number = {1},
  pages = {1 -- 8},
  year = {2006},
  doi = {10.1080/18811248.2006.9711062},
  url = {http://www.tandfonline.com/doi/abs/10.1080/18811248.2006.9711062}
}

@article{RIPL3,
  title = {{RIPL} - {Reference Input Parameter Library} for Calculation of Nuclear Reactions and Nuclear Data Evaluations},
  author = {Capote, R. and Herman, M. and Oblo\v{z}insk\'{y}, P.  and Young, P. G.  and Goriely, S.  and Belgya, T.  and Ignatyuk, A. V.  and Koning, A. J.  and Hilaire, S.  and Plujko, V. A.  and Avrigeanu, M.  and Bersillon, O.  and Chadwick, M. B.  and Fukahori, T.  and Ge, Z.  and Han, Y.  and Kailas, S.  and Kopecky, J.  and Maslov, V. M.  and Reffo, G.  and Sin, M.  and Soukhovitskii, E. Sh.  and Talou, P.},
  journal = {Nuclear Data Sheets},
  volume = {110},
  number = {12},
  pages = {3107 -- 3214},
  year = {2009},
  issn = {0090-3752},
  publisher = {Elsevier},
  doi = {10.1016/j.nds.2009.10.004},
  url = {http://www.sciencedirect.com/science/article/pii/S0090375209000994}
}

@article{Schiffer1963,
  title = {Search for a particle-stable tetra neutron},
  author = {Schiffer, J.P. and Vandenbosch, R.},
  journal = {Physics Letters},
  volume = {5},
  number = {4},
  pages = {292 -- 293},
  year = {1963},
  issn = {0031-9163},
  doi = {10.1016/S0375-9601(63)96134-6},
  url = {https://www.sciencedirect.com/science/article/pii/S0375960163961346},
}

@article{Bojowald1988,
  title = {Elastic deuteron scattering and optical model parameters at energies up to 100 {MeV}},
  author = {Bojowald, J. and Machner, H. and Nann, H. and Oelert, W. and Rogge, M. and Turek, P.},
  journal = {Phys. Rev. C},
  volume = {38},
  issue = {3},
  pages = {1153--1163},
  numpages = {0},
  year = {1988},
  month = {Sep},
  publisher = {American Physical Society},
  doi = {10.1103/PhysRevC.38.1153},
  url = {https://link.aps.org/doi/10.1103/PhysRevC.38.1153}
}

@techreport{Becchetti1969b,
  title = {Standard Optical Model Parameters},
  author = {Becchetti, F. D. and Greenlees, G. W.},
  journal = {1969 Annual Report, John H. Williams Laboratory of Nuclear Physics},
  pages = {116 -- 116},
  year = {1969},
  institution = {University of Minnesota},
  url = {https://www.osti.gov/servlets/purl/4908657/}
}

@article{Detraz1977,
  title = {Possible existence of bound neutral nuclei},
  author = {Detraz, Claude},
  journal = {Physics Letters B},
  volume = {66},
  number = {4},
  pages = {333 -- 336},
  year = {1977},
  issn = {0370-2693},
  doi = {10.1016/0370-2693(77)90008-9},
  url = {https://www.sciencedirect.com/science/article/pii/0370269377900089},
}

@article{Abdurrahman2024,
  title = {Neck rupture and scission neutrons in nuclear fission},
  author = {Abdurrahman, Ibrahim and Kafker, Matthew and Bulgac, Aurel and Stetcu Ionel},
  journal = {Phys. Rev. Lett.},
  volume = {132},
  number = {},
  pages = {242501},
  year = {2024},
  month = {Jun},
  publisher = {American Physical Society},
  doi = {10.1103/PhysRevLett.132.242501},
  url = {https://doi.org/10.1103/PhysRevLett.132.242501},
}

@article{Deltuva2018,
title = {Tetraneutron: Rigorous continuum calculation},
journal = {Physics Letters B},
volume = {782},
pages = {238-241},
year = {2018},
issn = {0370-2693},
doi = {https://doi.org/10.1016/j.physletb.2018.05.041},
url = {https://www.sciencedirect.com/science/article/pii/S0370269318304052},
author = {A. Deltuva}
}

@article{Higgins2021,
  title = {Comprehensive study of the three- and four-neutron systems at low energies},
  author = {Higgins, Michael D. and Greene, Chris H. and Kievsky, A. and Viviani, M.},
  journal = {Phys. Rev. C},
  volume = {103},
  issue = {2},
  pages = {024004},
  numpages = {12},
  year = {2021},
  month = {Feb},
  publisher = {American Physical Society},
  doi = {10.1103/PhysRevC.103.024004},
  url = {https://link.aps.org/doi/10.1103/PhysRevC.103.024004}
}

@article{Lazauskas2023,
  title = {Low Energy Structures in Nuclear Reactions with $4n$ in the Final State},
  author = {Lazauskas, Rimantas and Hiyama, Emiko and Carbonell, Jaume},
  journal = {Phys. Rev. Lett.},
  volume = {130},
  issue = {10},
  pages = {102501},
  numpages = {6},
  year = {2023},
  month = {Mar},
  publisher = {American Physical Society},
  doi = {10.1103/PhysRevLett.130.102501},
  url = {https://link.aps.org/doi/10.1103/PhysRevLett.130.102501}
}

@article{Grigorenko2004,
  title = {Broad states beyond the neutron drip line: Examples of {$^5\mathrm{H}$} and {$^4\mathrm{n}$}},
  volume = {19},
  ISSN = {1434-601X},
  url = {http://dx.doi.org/10.1140/epja/i2003-10124-1},
  DOI = {10.1140/epja/i2003-10124-1},
  number = {2},
  journal = {The European Physical Journal A},
  publisher = {Springer Science and Business Media LLC},
  author = {Grigorenko,  L. V. and Timofeyuk,  N. K. and Zhukov,  M. V.},
  year = {2004},
  month = Feb,
  pages = {187–201}
}

@article{Timofeyuk2003,
  title = {Do multineutrons exist?},
  volume = {29},
  ISSN = {0954-3899},
  url = {http://dx.doi.org/10.1088/0954-3899/29/2/102},
  DOI = {10.1088/0954-3899/29/2/102},
  number = {2},
  journal = {Journal of Physics G: Nuclear and Particle Physics},
  publisher = {IOP Publishing},
  author = {Timofeyuk,  N K},
  year = {2003},
  month = Jan,
  pages = {L9–L14}
}

@article{Pieper2003,
  title = {{Can Modern Nuclear Hamiltonians Tolerate a Bound Tetraneutron?}},
  volume = {90},
  ISSN = {1079-7114},
  url = {http://dx.doi.org/10.1103/PhysRevLett.90.252501},
  DOI = {10.1103/physrevlett.90.252501},
  number = {25},
  journal = {Physical Review Letters},
  publisher = {American Physical Society (APS)},
  author = {Pieper,  Steven C.},
  year = {2003},
  month = June 
}

@article{Bertulani2003,
  title = {Is the tetraneutron a bound dineutron–dineutron molecule?},
  volume = {29},
  ISSN = {1361-6471},
  url = {http://dx.doi.org/10.1088/0954-3899/29/10/309},
  DOI = {10.1088/0954-3899/29/10/309},
  number = {10},
  journal = {Journal of Physics G: Nuclear and Particle Physics},
  publisher = {IOP Publishing},
  author = {Bertulani,  C A and Zelevinsky,  V},
  year = {2003},
  month = Sept,
  pages = {2431–2437}
}

@article{Higgins2020,
  title = {Nonresonant Density of States Enhancement at Low Energies for Three or Four Neutrons},
  volume = {125},
  ISSN = {1079-7114},
  url = {http://dx.doi.org/10.1103/PhysRevLett.125.052501},
  DOI = {10.1103/physrevlett.125.052501},
  number = {5},
  journal = {Physical Review Letters},
  publisher = {American Physical Society (APS)},
  author = {Higgins,  Michael D. and Greene,  Chris H. and Kievsky,  A. and Viviani,  M.},
  year = {2020},
  month = July 
}

\end{document}